\documentclass[hyper]{JHEP} 

\usepackage{epsfig}

\newcommand\fverb{\setbox\pippobox=\hbox\bgroup\verb}

\newcommand\fverbdo{\egroup\medskip\noindent%

            \fbox{\unhbox\pippobox}\ }

\newcommand\fverbit{\egroup\item[\fbox{\unhbox\pippobox}]}

\newbox\pippobox

\title{Hamiltonian Formalism of Particular
Bimetric Gravity Model}
\author{J. Kluso\v{n}\\
Department of
Theoretical Physics and Astrophysics\\
Faculty of Science, Masaryk University\\
Kotl\'{a}\v{r}sk\'{a} 2, 611 37, Brno\\
Czech Republic\\
E-mail: \email{klu@physics.muni.cz}}
\preprint{}

 \abstract{In this short note we perform
 the  Hamiltonian analysis   of  bimetric gravity with one particular
 form of  potential between two metrics.
 We find that this theory have eight secondary constraints. We identify
 four constraints that are the first class constraints
 on condition
 when the interaction term  obeys some specific condition.
 We show that for the  form of the potential studied
 in this paper this condition is obeyed and hence
 we can interpret these first class constraints as generators
 of the diagonal diffeomorphism.}

\def\mR{\mathcal{R}}

\def\hf{\hat{f}}
\def\tK{\tilde{K}}

\def\bN{\bar{N}}

\def\bG{\mathbf{G}}
\def\bH{\mathbf{H}}

\def\bN{\bar{N}}

\def\tnabla{\tilde{\nabla}}

\def\tmG{\tilde{\mG}}

\def\be{\begin{equation}}

\def\ee{\end{equation}}

\def\bea{\begin{eqnarray}}

\def\eea{\end{eqnarray}}

\def\bG{\mathbf{G}}

\def\bx{\mathbf{x}}
\def\by{\mathbf{y}}

\newcommand{\hg}{\hat{g}}

\newcommand{\mG}{\mathcal{G}}

\def\mV{\mathcal{V}}

\newcommand{\bT}{\mathbf{T}}
\def\bmR{\mathbf{\mR}}

\def\pb #1{\left\{#1\right\}}

\begin{document}
\section{Introduction and Summary}
Bimetric theories of  gravity are basically defined as theories
of two metrics $\hg_{\mu\nu}$ and $\hf_{\mu\nu}$ whose
 dynamics are governed by two Einstein-Hilbert
actions together with the  interaction term between $\hg_{\mu\nu}$
and $\hf_{\mu\nu}$. The history of bimetric or more generally
multimetric theories of gravity \cite{Hinterbichler:2012cn} is long,
we recommend the paper \cite{Damour:2002ws} for list of the
references to earlier works \footnote{For recent works, see
\cite{Kuhnel:2012gh,Hassan:2012wr,Deffayet:2012nr,Nomura:2012xr,
Hassan:2012wt,vonStrauss:2011mq,Hassan:2011zd,Banados:2011hk,Banados:2009it,
Banados:2008fi}.}. It was believed for the long time that these
theories suffer from the presence of the ghost degree of freedom.
However very recently the ghost free bimetric theory of gravity was
suggested in \cite{Hassan:2011zd}. The main novelty of this new
bimetric theory of the gravity is  the specific form of the
interaction term between $\hg_{\mu\nu}$ and $\hf_{\mu\nu}$ that was
firstly proposed in the formulation of the ghost-free non-linear
massive theory of gravity
\cite{deRham:2010kj,Hassan:2011vm,Hassan:2011hr,deRham:2011rn}.

It is well known that  the bimetric theories of gravity are
invariant under diagonal  diffeomorphism. More explicitly, without
the interaction term the action for two bimetric theory of gravity
is sum of two Einstein-Hilbert actions for $\hg_{\mu\nu}$ and
$\hf_{\mu\nu}$ and each of these actions is diffeomorphism
invariant. However the  presence of the interaction term breaks this
symmetry to the diagonal one. Since this is the gauge symmetry we
should expect an existence of the four first class constraints in
the Hamiltonian formalism of given theory. Remarkably  it turns out
that it is non-trivial task to find these constraints and to show
that they are really the first class constraints.
 This short note is
devoted to this analysis at least for same specific form of the
potential term.

Explicitly, we perform  the $3+1$ splitting of the metric components
$\hg_{\mu\nu},\hf_{\mu\nu}$ and determine corresponding Hamiltonian.
Then, following seminar paper \cite{Damour:2002ws} we perform the
redefinition of the lapse and shifts functions so that the
Hamiltonian is linear in the new lapse function and the new shift
functions. Clearly this is the essential condition for the existence
of the diffeomorphism constraints however it is not sufficient. In
fact, we have to show that these constraints are preserved during
the time evolution of the system without imposing  conditions on the
Lagrange multipliers. In other words we have to show that the
Poisson brackets between diffeomorphism constraints vanish on the
constraint surface. We show that the right side of the Poisson
bracket between Hamiltonian constraints is proportional to the
particular linear combinations of the variation of the potential
term and we show that this expression vanishes for the case of the
potential studied in this paper. In other words we find that there
exists the Hamiltonian constraint and three spatial diffeomorphism
constraints and that the Poisson brackets between these constraints
vanish on the constraint surface. We believe that this is
non-trivial result since as far as we know such an analysis has not
been performed yet. Then we also analyze the remaining constraints
and we show that they are the second class constraints. Finally we
determine the number of the physical degrees of freedom and we argue
that they can be interpreted as  massless graviton, massive graviton
and one additional scalar mode at linearized level.

It is clear that our work has an important limitation in the special
form of the potential that was chosen. The natural extension of this
work   is to extend given analysis to the case of the bimetric
theories of gravity introduced in \cite{Hassan:2011zd}. However even
if it was argued there that these theories are ghost free it is not
completely clear how to identify the generators of the diagonal
diffeomorphism. In principle the analysis presented here could be
applied to this case as well but it is not clear how to identify
additional constraints in given theory that are crucial for the
elimination of the scalar mode. This problem is currently under
investigation.
\section{Hamiltonian analysis of  Bimetric gravity}
\label{second} We begin this section with the introduction of the
bimetric theories of gravity. The basic idea of bimetric gravity is
simple. We have two Einstein-Hilbert actions for two four
dimensional metrics $\hg_{\mu\nu},\hf_{\mu\nu} \ , \mu,\nu=0,1,2,3$
together with the interaction term that  does not contain the
derivatives of the metric \footnote{We follow the notation used in
\cite{Damour:2002ws}.}
\begin{eqnarray}\label{bigraction}
S&=&M_L^2\int d^4x \sqrt{-\hg}{}^{(4)} R(\hg)+M_R^2 \int
d^4x\sqrt{-\hf} {}^{(4)}R(\hf)-\nonumber \\
&-&\mu\int d^4x (\det \hg\det\hf)^{1/4} \mV(\hg,\hf) \ . \nonumber \\
\end{eqnarray}
In this short note we  consider following specific form of the
potential $\mV$
\begin{equation}\label{defPot}
\mV(\hg,\hf)=\sum_nc_n(H^\mu_{ \ \mu})^n+ \sum_m d_m (H^\mu_{ \
\nu}H^\nu_{ \ \mu})^m \ ,
\end{equation}
where \begin{equation} H^\mu_{ \ \nu}=\hg^{\mu\rho} \hf_{\rho\nu} \
,
\end{equation}
and where $c_n$ and $d_n$ are numerical constants.  Note that  the
action (\ref{bigraction}) is invariant under following
diffeomorphism transformations
\begin{equation}\label{diffov}
\hg'_{\mu\nu}(x')=
 \hg_{\rho\sigma}(x) \frac{\partial x^\rho}{\partial x'^\mu}
\frac{\partial x^\sigma}{\partial x'^\nu}
 \ ,
 \quad
\hf'_{\mu\nu}(x')= \hf_{\rho\sigma}(x) \frac{\partial
x^\rho}{\partial x'^\mu} \frac{\partial x^\sigma}{\partial x'^\nu} \
.
\end{equation}
Our goal is to perform the Hamiltonian analysis of the theory
defined by the action (\ref{bigraction}).

To begin with we introduce the $3+1$ decomposition of the four
dimensional metric $\hat{g}_{\mu\nu}$
\cite{Gourgoulhon:2007ue,Arnowitt:1962hi}
\begin{eqnarray}
\hat{g}_{00}=-N^2+N_i g^{ij}N_j \ , \quad \hat{g}_{0i}=N_i \ , \quad
\hat{g}_{ij}=g_{ij} \ ,
\nonumber \\
\hat{g}^{00}=-\frac{1}{N^2} \ , \quad \hat{g}^{0i}=\frac{N^i}{N^2} \
, \quad \hat{g}^{ij}=g^{ij}-\frac{N^i N^j}{N^2} \
\nonumber \\
\end{eqnarray}
together with the metric $\hf_{\mu\nu}$
\begin{eqnarray}
\hf_{00}&=&-M^2+L_i f^{ij}L_j \ , \quad \hf_{0i}=L_i \ , \quad
\hf_{ij}=f_{ij} \ , \nonumber \\
\hf^{00}&=&-\frac{1}{M^2} \ , \quad  \hf^{0i}=\frac{L^i}{M^2} \ ,
\quad \hf^{ij}=
f^{ij}-\frac{L^i L^j}{M^2} \ , \quad  L^i=L_jf^{ji} \ . \nonumber \\
\end{eqnarray}
Then using the well known relation \footnote{We ignore the boundary
terms.}
\begin{eqnarray}\label{Rfour}
{}^{(4)}R[\hg]&=&K_{ij}\mG^{ijkl}K_{kl}+R^{(g)} \ , \nonumber \\
{}^{(4)}R[\hf]&=&\tK_{ij}\tmG^{ijkl}\tK_{kl}+R^{(f)} \ , \nonumber \\
\end{eqnarray}
where $R^{(g)}$ and $R^{(f)}$ are three dimensional scalar
curvatures evaluated using the spatial metric $g_{ij}$ and $f_{ij}$
respectively and  where the extrinsic curvatures $K_{ij}$ and
$\tK_{ij}$ are defined as
\begin{equation}
K_{ij}=\frac{1}{2N}(\partial_t g_{ij}- \nabla_i N_j-\nabla_j N_i)\ ,
\quad
 \tK_{ij}=\frac{1}{2M}(\partial_t f_{ij}- \tnabla_i
L_j-\tnabla_j L_i) \ ,
\end{equation}
and where $\nabla_i$ and $\tnabla_i$ are covariant derivatives
evaluated using the metric components $g_{ij}$ and $f_{ij}$
respectively. Finally note that $\mG^{ijkl}$ and $\tmG^{ijkl}$ are
de Witt metrics defined as
\begin{equation}
\mG^{ijkl}=\frac{1}{2}(g^{ik}g^{jl}+g^{il}g^{jk})-g^{ij}g^{kl} \ ,
\quad  \tmG^{ijkl}=\frac{1}{2}(f^{ik}f^{jl}+f^{il}f^{jk})-
f^{ij}f^{kl} \
\end{equation}
with inverse
\begin{equation}
\mG_{ijkl}=\frac{1}{2}(g_{ik}g_{jl}+ g_{il}g_{jk})-\frac{1}{2}
g_{ij}g_{kl} \ , \quad \tmG_{ijkl}=\frac{1}{2}(f_{ik}f_{jl}+
f_{il}f_{jk})-\frac{1}{2} f_{ij}f_{kl} \
\end{equation}
that obey the relation
\begin{equation}
\mG_{ijkl}\mG^{klmn}=\frac{1}{2}(\delta_i^m\delta_j^n+
\delta_i^n\delta_j^m)  \ , \quad
\tmG_{ijkl}\tmG^{klmn}=\frac{1}{2}(\delta_i^m\delta_j^n+
\delta_i^n\delta_j^m)  \ .
\end{equation}
Using (\ref{Rfour}) we rewrite the action (\ref{bigraction}) into
the form that is suitable for the Hamiltonian analysis
\begin{eqnarray}\label{SFRbi}
S&=&\int dt L=M_g^2 \int d^3\bx dt
\sqrt{g}N[K_{ij}\mG^{ijkl}K_{kl}+R^{(g)}
]+ \nonumber \\
&+& M_f^2\int d^3 \bx dt M\sqrt{f} [\tK_{ij}\tmG^{ijkl}\tK_{kl}+
R^{(f)}]-\mu \int d^3\bx dt g^{1/4}f^{1/4}\sqrt{NM}\mV \ .
 \nonumber \\
\end{eqnarray}
Then  from (\ref{SFRbi}) we determine following conjugate momenta
\begin{eqnarray}
\pi^{ij}&=&\frac{\delta L}{\delta \partial_t g_{ij}}=
M_g^2\mG^{ijkl}K_{kl} \ , \quad \rho^{ij}= \frac{\delta L}{\delta
\partial_t f_{ij}}= M_f^2\tmG^{ijkl}\tK_{kl} \ ,
\nonumber \\
\pi_i&=&\frac{\delta L}{\delta \partial_t N^i}\approx 0 \ , \quad
\rho_i=\frac{\delta L}{\delta \partial_t L^i}\approx 0 \ , \nonumber
\\
\pi_N&=&\frac{\delta L}{\delta \partial_t N}\approx 0 \ , \quad
\rho_M=\frac{\delta L}{\delta \partial_t M}\approx 0 \  \nonumber
\\
\end{eqnarray}
and then using the  standard procedure we derive following
Hamiltonian
\begin{eqnarray}
H=\int d^3\bx (N \mR_0^{(g)}+M \mR_0^{(f)}+ N^i\mR_i^{(g)}+
L^i\mR_i^{(f)}+
\mu \sqrt{NM}g^{1/4}f^{1/4}\mV) \  ,  \nonumber \\
\end{eqnarray}
where
\begin{eqnarray}
\mR_0^{(g)}&=&\frac{1}{M_g^2\sqrt{g}}
\pi^{ij}\mG_{ijkl}\pi^{kl}-M_g^2\sqrt{g}R^{(g)} \ , \quad
\mR_0^{(f)}= \frac{1}{M_f^2
\sqrt{f}}\rho^{ij}\tmG_{ijkl}\rho^{kl}-M_f^2\sqrt{f}R^{(f)} \ , \nonumber \\
\mR_i^{(g)}&=&-2g_{ij}\nabla_k\pi^{kj} \ , \quad \mR_i^{(f)}=
-2f_{ij}\tnabla_k\rho^{kj} \ . \nonumber \\
\end{eqnarray}
The crucial point is to identify four constraints that correspond to
the diffeomorphism invariance of given theory. In order to do this
we proceed as in \cite{Damour:2002ws} and introduce following
variables
\begin{eqnarray}\label{defnewN}
\bN&=&\sqrt{NM} \ , \quad  n=\sqrt{\frac{N}{M}} \ , \quad
\bN^i=\frac{1}{2}
(N^i+L^i) \ , \quad  n^i=\frac{N^i-L^i}{\sqrt{NM}} \ , \nonumber \\
N&=&\bN n \ , \quad  M=\frac{\bN}{n} \ , \quad
L^i=\bN^i-\frac{1}{2}n^i\bN \ , \quad  N^i=\bN^i+\frac{1}{2}n^i\bN
 \ , \nonumber \\
\end{eqnarray}
where again clearly their conjugate momenta are the primary
constraints of the theory
\begin{equation}\label{Pbnprim}
P_{\bN}\approx 0 \ , \quad  P_n\approx 0 \ , \quad  P_i\approx 0 \ ,
\quad  p_i\approx 0 \ .
\end{equation}
Note that the canonical variables have following  non-zero Poisson
brackets
\begin{eqnarray}\label{canpb}
\pb{g_{ij}(\bx),\pi^{kl}(\by)}&=& \frac{1}{2}(\delta_i^k\delta_j^l+
\delta_j^k\delta_i^l)\delta(\bx-\by) \ , \quad
\pb{f_{ij}(\bx),\rho^{kl}(\by)}= \frac{1}{2}(\delta_i^k\delta_j^l+
\delta_j^k\delta_i^l)\delta(\bx-\by) \ , \nonumber \\
\pb{\bN(\bx),P_{\bN}(\by)}&=&\delta(\bx-\by) \ , \quad
 \pb{n(\bx),P_{n}(\by)}=\delta(\bx-\by) \ , \nonumber \\
\pb{\bN^i(\bx),P_j(\by)}&=&\delta^i_j\delta(\bx-\by) \ , \quad
\pb{n^i(\bx),p_j(\by)}=\delta^i_{j}\delta(\bx-\by) \ .
\nonumber \\
\end{eqnarray}
With the help of (\ref{defnewN}) we find  the explicit form of the matrix
$H^\mu_{ \ \nu}$
\begin{eqnarray}
H^0_{ \ 0}
&=&\frac{1}{n^2} +\frac{n^i}{\bN n^2}f_{ij}(\bN^j-\frac{1}{2}\bN
n^j)
 \ , \quad
H^0_{ \ j}
= \frac{1}{\bN n^2}n^kf_{kj} \  ,
 \nonumber \\
H^i_{ \ 0}&=&
-\frac{1}{n^4}(\bN^i+\frac{1}{2}\bN n^i)
 +g^{ij}f_{jk}(\bN^k-\frac{1}{2}n^k\bN)
 -\frac{1}{\bN n^2}(\bN^i+\frac{1}{2}n^i\bN)
 n^kf_{kl}(\bN^l-\frac{1}{2}\bN n^l) \ , \nonumber \\
H^i_{ \ j}&=&
 g^{ik}f_{kj}-\frac{1}{\bN n^2}(\bN^i+\frac{1}{2}n^i\bN)n^kf_{kj}
\nonumber \\
\end{eqnarray}
so that
\begin{equation}
H^\mu_{ \ \mu}=\frac{1}{n^4}-\frac{1}{n^2}n^if_{ij}f^j+g^{ij}f_{ji}
\end{equation}
and also
\begin{eqnarray}
 H^\mu_{ \ \nu}H^\nu_{ \ \mu}=
\frac{1}{n^8}-\frac{2}{n^6}n^if_{ij}n^j+\frac{1}{n^4}(n^if_{ij}n^j)^2+
g^{ij}f_{jk}g^{kl}f_{li}-\frac{2}{n^2} n^if_{ij}g^{jk}f_{kl}n^l \ .
\nonumber \\
\end{eqnarray}
As a result we find  that the potential $\mV$ defined in
(\ref{defPot}) does not depend on $\bN,\bN^i$ which is very
important for the  existence of the
  diffeomorphism constraints.
Then  using the variables (\ref{defnewN}) we find that the
Hamiltonian takes the form
\begin{equation}\label{defHnew}
H=\int d^3\bx (\bN \bmR+\bN^i\bmR_i) \ ,
\end{equation}
where
\begin{eqnarray}
\bmR&=&n\mR_0^{(g)}+\frac{1}{n}\mR_0^{(f)}+
\frac{1}{2}n^i\mR_i^{(g)}
-\frac{1}{2}n^i\mR_i^{(f)}+\nonumber \\
&+&\mu  g^{1/4}f^{1/4}\mV \  , \quad  \bmR_i=
\mR_i^{(g)}+\mR_i^{(f)} \ . \nonumber \\
\end{eqnarray}
As usual  the requirement of the preservation of the primary
constraints (\ref{Pbnprim})  implies following secondary ones
\begin{eqnarray}
\partial_t P_{\bN}&=&\pb{P_{\bN},H}=-\bmR\approx 0 \ , \nonumber \\
\partial_t P_i&=&\pb{P_i,H}=-\bmR_i \approx 0 \ , \nonumber \\
\partial_t P_n&=&\pb{P_n,H}=-\mR_0^{(g)}+
\frac{1}{n^2}\mR_0^{(f)}-\mu  g^{1/4}f^{1/4}\frac{\delta
\mV}{\delta n}\equiv \mG\approx 0 \ , \nonumber \\
\partial_t p_i&=&\pb{p_i,H}=-\frac{1}{2}\mR_i^{(g)}+
\frac{1}{2}\mR_i^{(f)}-\mu  g^{1/4}f^{1/4}\frac{\delta
\mV}{\delta n^i}\equiv \mG_i\approx 0 \ . \nonumber \\
\end{eqnarray}
For the consistency of the theory it is important to show that the
constraints $\mR$ and $\mR_i$ are the first class constraints.
 To proceed  it is useful to introduce the smeared form of
 the constraint $\mR$
\begin{equation}\label{defsmearconsH}
\bT_T(N)=\int d^3\bx N(\bx)\mR(\bx) \ .
\end{equation}
In case of the constraint $\mR_i$ it turns out that it is convenient
to extend the constraint $\mR_i$ with appropriate combinations of
the primary constraints $P_n,p_i$ so that we define $\tilde{\mR}_i$
as
\begin{equation}
\tilde{\mR}_i=\mR_i^{(g)}+\mR_i^{(f)}+\partial_inP_n+\partial_in^jp_j+
\partial_j (n^j p_i) \
\end{equation}
and then define its smeared form
\begin{equation}\label{defsmearconsHs}
\bT_S(N^i)=\int d^3\bx N^i\tilde{\mR}_i \ .
\end{equation}
Finally it is useful to introduce
 the smeared forms of the constraints $\mR_0^{(f),(g)}$
and $\mR_i^{(f),(g)}$
\begin{eqnarray}
 \bT_T^g(N)&=&\int d^3\bx N(\bx)\mR_0^{(g)}(\bx) \ , \quad
  \bT_T^f(N)= \int
d^3\bx N(\bx)\mR_0^{(f)}(\bx) \ , \nonumber \\
 \bT_S^g(N^i)&=&\int
d^3\bx N^i(\bx)\mR_i^{(g)}(\bx)
\ , \quad
 \bT_S^f(N^i)=\int d^3\bx N^i(\bx)\mR_i^{(f)}(\bx) \ . \nonumber \\
\end{eqnarray}
It is well known that these smeared constraints have following
non-zero Poisson brackets \footnote{See, for example
\cite{Hojman:1976vp}.}
\begin{eqnarray}\label{pbGR}
\pb{\bT^g_T(N),\bT^g_T(M)}&=& \bT^g_S((N\partial_iM-M\partial_i
N)g^{ij}) \
, \nonumber \\
\pb{\bT^f_T(N),\bT^f_T(M)}&=& \bT^f_S((N\partial_iM-M\partial_i
N)f^{ij}) \
, \nonumber \\
\pb{\bT_S^g(N^i),\bT_T^g(M)}&=&\bT_T^g(N^i\partial_i M) \ , \nonumber \\
\pb{\bT_S^f(N^i),\bT_T^f(M)}&=&\bT_T^f(N^i\partial_i M) \ ,  \nonumber
\\
\pb{\bT_S^g(N^i),\bT_S^g(M^j)}&=&\bT^g_S((N^j\partial_j
M^i-M^j\partial_j N^i)) \ , \nonumber \\
\pb{\bT_S^f(N^i),\bT_S^f(M^j)}&=&\bT^f_S((N^j\partial_j
M^i-M^j\partial_j N^i)) \ . \nonumber \\
\end{eqnarray}
To proceed further note that using (\ref{canpb}) and
(\ref{defsmearconsHs}) we find
\begin{eqnarray}\label{bTScan}
\pb{\bT_S(N^i),g_{ij}}&=&-N^k\partial_k g_{ij}-\partial_i N^kg_{kj}-
g_{ik}\partial_jN^k \ ,  \nonumber \\
\pb{\bT_S(N^i),\pi^{ij}}&=& -\partial_k(N^k\pi^{ij})
+\partial_kN^i\pi^{kj}+\pi^{ik}\partial_k N^j \ , \nonumber \\
\pb{\bT_S(N^i),f_{ij}}&=& -N^k\partial_k f_{ij}-
\partial_iN^kf_{kj}-f_{ik}\partial_jN^k \ , \nonumber \\
\pb{\bT_S(N^i),\rho^{ij}}&=&-\partial_k (N^k\rho^{ij}) +\partial_k
N^i\rho^{kj}+\rho^{ik}\partial_k N^j \ , \nonumber \\
\pb{\bT_S(N^i),n}&=&-N^i\partial_i n \ , \nonumber \\
\pb{\bT_S(N^i),P_n}&=&-\partial_i (N^iP_n) \ , \nonumber \\
\pb{\bT_S(N^i),n^i}&=&-N^k\partial_k n^i+\partial_j N^in^j \ ,
\nonumber \\
\pb{\bT_S(N^i),p_i}&=&-\partial_k (N^kp_i)-\partial_i N^kp_k \ .
\nonumber \\
\end{eqnarray}
From these results we  see that $\bT_S(N)$ is the generator of the
spatial diffeomorphism. Moreover, using previous Poisson brackets we
easily find  that
\begin{equation}\label{pbbTSS}
\pb{\bT_S(N^i),\bT_S(M^j)}= \bT_S((N^j\partial_j M^i-M^j\partial_j
N^i)) \ .
\end{equation}
On the other hand more interesting is to determine the Poisson
bracket between smeared forms of the Hamiltonian constrains
(\ref{defsmearconsH})
\begin{eqnarray}\label{pbbTT}
& &\pb{\bT_T(N),\bT_T(M)}= \pb{\bT^g_T(nN),\bT^g_T(nM)} +
\pb{\bT^f_T(\frac{1}{n}N),\bT^f_T(\frac{1}{n}M)}+\nonumber
\\
&+&\pb{\bT^{g}_T(Nn), \bT_S^{g}(\frac{1}{2}M n^i)}
+\pb{\bT^{g}_S(\frac{1}{2}Nn^i),\bT_T^{g}(Mn)}+
 \nonumber \\
 &+&\pb{\bT^{g}_T(nN),\int d^3\bx N \mu  g^{1/4}f^{1/4}
 \mV}+\pb{\int d^3\bx M\mu  g^{1/4}f^{1/4}\mV,\bT^{g}_T(nM)}+
 \nonumber \\
&-&\pb{\bT^{f}_T(N\frac{1}{n}), \bT_S^{f}(\frac{1}{2}M n^i)}
-\pb{\bT^{f}_S(\frac{1}{2}Nn^i),\bT_T^{f}(M\frac{1}{n})}+
 \nonumber \\
 &+&\pb{\bT^{f}_T(\frac{1}{n}N),\int d^3\bx N \mu  g^{1/4}f^{1/4}
 \mV}+\pb{\int d^3\bx M\mu  g^{1/4}f^{1/4}\mV,\bT^{f}_T(\frac{1}{n}M)}+
 \nonumber \\
 &+&\frac{1}{4}\pb{\bT^{g}_S(Nn^i),\bT^{g}_S(Mn^j)}+
 \frac{1}{4}\pb{\bT^{f}_S(Nn^i),\bT^{f}_S(Mn^j)}+
 \nonumber \\
 &+&\frac{1}{2}\pb{\bT^{g}_S(Nn^i),\int d^3\bx M\mu f^{1/4}
 g^{1/4}\mV}+
 \frac{1}{2}\pb{\int d^3\bx \mu N f^{1/4}g^{1/4}
 \mV,\bT^{g}_S(Mn^i)}-\nonumber \\
&-&\frac{1}{2}\pb{\bT^{f}_S(Nn^i),\int d^3\bx M\mu f^{1/4}
 g^{1/4}\mV}-
 \frac{1}{2}\pb{\int d^3\bx \mu N f^{1/4}g^{1/4}
 \mV,\bT^{f}_S(Mn^i)}=\nonumber \\
 &=&\bT_S((N\partial_i M-M\partial_i N)n^2 g^{ij})+
\bT_S((N\partial_i M-M\partial_i N)\frac{1}{n^2}f^{ij})
 -\nonumber \\
 &-&\bG_S((N\partial_i M-M\partial_i N)n^2 g^{ij})
-\bG_T((N\partial_iM-M\partial_iN)n^i)
+\nonumber \\
&+&\bG_S((N\partial_i M-M\partial_iN)\frac{1}{n^2}f^{ij}) +\int
d^3\bx
(N\partial_i M-M\partial_i N)\Sigma^i \ , \nonumber \\
\end{eqnarray}
where we defined the smeared forms of the constraints $\mG_i$ and
$\mG$
\begin{equation}
\bG_T(N)=\int d^3\bx N(\bx) \mG(\bx)
 \ , \quad  \bG_S(N^i)= \int d^3\bx
N^i(\bx)\mG_i(\bx) \ ,
\end{equation}
and where $\Sigma^i$ is defined as
\begin{eqnarray}
\Sigma^i=\mu  g^{1/4}f^{1/4}\left[-n^2g^{ij}
 \frac{\delta \mV}{\delta n^j}
 +f^{ij}\frac{\delta \mV}{\delta n^j}\frac{1}{n^2}
- \frac{\delta \mV}{\delta g^{kj}}n^kg^{ij}- \frac{\delta
\mV}{\delta f_{ij}} n^kf_{kj} -\frac{1}{2}n^in\frac{\delta
\mV}{\delta n} \right] \ .  \nonumber \\
\end{eqnarray}
We see that the Poisson bracket between the smeared forms of the
Hamiltonian constraints (\ref{pbbTT}) vanish on the constraint
surface on condition that $\Sigma_i$ is zero. We  explicitly check
below  that this is indeed the case for the potential
(\ref{defPot}).
In fact, using
\begin{eqnarray}\label{varH}
\frac{\delta H^\mu_{ \ \mu}}{\delta
n}&=&-\frac{4}{n^5}+\frac{2n^if_{ij}n^j}{n^3} \ , \quad \frac{\delta
H^\mu_{ \ \mu}}{\delta n^i}= -2\frac{f_{ij}n^j}{n^2} \ , \nonumber
\\
\frac{\delta H^\mu_{ \ \mu}}{\delta g^{ij}}&=& f_{ij} \ , \quad
\frac{\delta
 H^\mu_{ \ \mu}}{\delta
f_{ij}}=g^{ij}-\frac{n^i n^j}{n^2} \  \nonumber \\
\end{eqnarray}
and
%
\begin{eqnarray}\label{varHH}
\frac{\delta (H^\mu_{ \ \nu}H^\nu_{ \ \mu})}{\delta
n}&=&-\frac{8}{n^9}+ 12 \frac{n^if_{ij}n^j}{n^7}-
\frac{4}{n^5}(n^if_{ij}n^j)^2+ \frac{4}{n^3}
n^mf_{mi}g^{ij}f_{jk}n^k
 \ ,
\nonumber \\
\frac{\delta (H^\mu_{ \ \nu}H^\nu_{ \ \mu})}{\delta
n^i}&=&-4\frac{f_{ij}n^j}{n^6} +\frac{4}{n^4}
f_{ij}n^j(n^kf_{kl}n^l)-\frac{4}{n^2}f_{ij}g^{jm}f_{mn}n^n \ ,
 \nonumber \\
\frac{\delta (H^\mu_{ \ \nu}H^\nu_{ \ \mu})}{\delta f_{ij}}&=&
-2\frac{n^in^j}{n^6} +\frac{2}{n^4} n^in^j(n^kf_{kl}n^l)+2
g^{im}f_{mn}g^{nj} -\frac{4}{n^2}g^{im}f_{mn}n^nn^j \ ,
 \nonumber \\
\frac{\delta (H^\mu_{ \ \nu}H^\nu_{ \ \mu})}{\delta g^{ij}}&=&
2f_{im}g^{mn}f_{nj} -\frac{2}{n^2} n^mf_{mi}f_{jn}n^n \
 \nonumber \\
\end{eqnarray}
and after some tedious calculations
 we  find that $\Sigma^i=0$ for the
potential (\ref{defPot}). In summary, we derived the fundamental
result that the Poisson bracket between Hamiltonian constraint
(\ref{pbbTT}) vanishes on the constraint surface.

 As the next step we determine the Poisson bracket
between $\bT_S(N^i)$ and $\bT_T(N)$. We firstly determine following
Poisson bracket
\begin{eqnarray}
\pb{\bT_S(N),g^{1/4}f^{1/4}\mV}&=&
-N^k\partial_k[g^{1/4}f^{1/4}\mV]-\partial_k N^k g^{1/4}f^{1/4}\mV+
\nonumber \\
&+&g^{1/4}f^{1/4}\left[ \frac{\delta \mV}{\delta n^i}\partial_j
N^in^j- 2\frac{\delta \mV}{\delta f_{kl}}\partial_k N^m f_{ml}+
2\frac{\delta \mV}{\delta g^{kl}}\partial_m N^k g^{ml}\right]=\nonumber \\
&=&-N^k\partial_k[g^{1/4}f^{1/4}\mV]-\partial_k N^k
g^{1/4}f^{1/4}\mV
\nonumber \\
\end{eqnarray}
where in the final step we used (\ref{varH}) and (\ref{varHH}). Then
with the help of  (\ref{pbGR}) and  (\ref{bTScan})
 we obtain
\begin{eqnarray}
\pb{\bT_S(N^i),\bT_T(M)}=\bT_T^{g}
(N^i\partial_i M n)+\bT_T^{f}(N^i\partial_i M\frac{1}{n^2})+ \nonumber \\
+\frac{1}{2}\bT_S^{g}(N^j\partial_j M n^i)- \frac{1}{2}\bT_S^{f}
(N^j\partial_j M n^i)+\nonumber \\
+\int d^3\bx  N^k\partial_k M\mu g^{1/4}f^{1/4}
\mV=\bT_T(N^i\partial_i M) \ .\nonumber \\
\end{eqnarray}
This result together with (\ref{pbbTSS}) and (\ref{pbbTT}) shows
that $\bT_T(N)$ and $\bT_S(N^i)$ are the first class constraints
that are generators of the diagonal  diffeomorphism.

Now we proceed  to the analysis of constraints $\mG_i,\mG_n$. For
further purposes we introduce following "Hamiltonian"
\begin{equation}
\bH(\bN,\bN^i)=\bT_T(\bN)+\bT_S(\bN^i) \
\end{equation}
so that the total Hamiltonian has the form
\begin{equation}
H_T=\bH(\bN,\bN^i)+ \int d^3\bx (v^n P_n+v^i p_i+u^n\mG_n+u^i \mG_i)
\ .
\end{equation}
Then the requirement of the preservation of the primary constraints
$P_n,p_i$ implies
\begin{eqnarray}\label{timePn}
\partial_t P_n&=&\pb{P_n,H_T}=\mG+
\int d^3\bx (u^n(\bx)\pb{P_n,\mG_n(\bx)}+ u^i\pb{P_n,\mG_i(\bx)})
\approx
\nonumber \\
&\approx& \int d^3\bx (u^n(\bx)\pb{P_n,\mG_n(\bx)}+
u^i\pb{P_n,\mG_i(\bx)})=0
\nonumber \\
\partial_t p_i &=&\pb{p_i,H_T}=\mG_i+
\int d^3\bx (u^n(\bx)\pb{p_i,\mG_n(\bx)}+ u^j\pb{p_i,\mG_j(\bx)})
\approx 0
\nonumber \\
&\approx &\int d^3\bx (u^n(\bx)\pb{p_i,\mG_n(\bx)}+
u^j\pb{p_i,\mG_j(\bx)}) = 0 \ .
\end{eqnarray}
We have four equations for four unknown $u_n$ and $u_i$ that can be
solved for $u_n,u_i$ at least in principle. Further,  the
requirement of the preservation of the constraints $\mG_n,\mG_i$
gives next four equations for unknown $v_n$ and $v_i$ that can again
be explicitly solved \footnote{However we should stress that there
is a possibility that with the suitable  form of the potential there
could exist   additional constraints. Such a form of the potential
is well known in the case of  the non-linear massive gravity case
and corresponding bi-metric generalization
\cite{Hinterbichler:2012cn,Hassan:2011zd}.}. In other words
$P_n,p_i,\mG_n,\mG_i$ are the second class constraints.

 For example, let us consider the
simplest form of the potential
\begin{equation}
\mV=H^\mu_{ \ \mu} \ .
\end{equation}
In this case we find
\begin{eqnarray}
\mG_n&=&-\mR_0^{(g)}+\frac{1}{n^2}\mR^{(f)}+\mu
f^{1/4}g^{1/4}\left(-\frac{4}{n^5}+2\frac{n^if_{ij}n^j}{n^3}\right)
 \ , \nonumber \\
\mG_i&=&-\frac{1}{2}\mR_i^{(g)}+ \frac{1}{2}\mR_i^{(f)}+2\mu
\frac{g^{1/4}f^{1/4}}{n^2}
f_{ij}n^j \nonumber \\
\end{eqnarray}
and hence
\begin{eqnarray}
\pb{P_n(\bx),\mG_n(\by)}&=&\left(\frac{2}{n^3}\mR_0^{(f)}
-20f^{1/4}g^{1/4}\frac{\mu}{n^6}+6\mu
f^{1/4}g^{1/4}\frac{n^if_{ij}n^j}{n^4}\right)\delta(\bx-\by)\nonumber
\\
&\equiv & \triangle_{nn}(\bx)\delta(\bx-\by) \ , \nonumber \\
\pb{P_n(\bx),\mG_i(\by)}&=&\frac{4\mu}{n^4}
g^{1/4}f^{1/4}f_{ij}n^j(\bx)\delta(\bx-\by)\equiv
\triangle_{ni}(\bx)
\delta(\bx-\by) \ , \nonumber \\
\pb{p_i(\bx),\mG_n(\by)}&=&-4\mu
f^{1/4}g^{1/4}\frac{f_{ij}n^j}{n^3}(\bx)\delta(\bx-\by) \equiv
\triangle_{in}(\bx)\delta(\bx-\by) \ ,
\nonumber \\
\pb{p_i(\bx),\mG_j(\by)}&=&-2\mu g^{1/4}f^{1/4}\frac{f_{ij}}{n^2}
(\bx)\delta(\bx-\by)\equiv \triangle_{ij}(\bx)\delta(\bx-\by) \ .
\nonumber \\
\end{eqnarray}
Then the first equation in (\ref{timePn}) can be solved for $u_n$ as
\begin{equation}\label{unui}
u^n=-\frac{\triangle_{ni}u^i}{\triangle_{nn}} \ ,
\end{equation}
Note that $\triangle_{nn}$ is non-zero for the generic point of the
phase space. Inserting this result into the second equation in
(\ref{timePn}) we obtain the homogeneous equation for $u^i$
\begin{equation}\label{equi}
\left(\triangle_{ij}-\frac{\triangle_{in}\triangle_{nj}}
{\triangle_{nn}}\right)u^j= \triangle_{im}\left(\delta^m_{ \ j}
-(\triangle^{-1})^{mk}
\frac{\triangle_{kn}\triangle_{nj}}{\triangle_{nn}}\right)u^j\equiv
\triangle_{im}M^m_{ \ j}u^j=0 \ .
\end{equation}
Now it is easy to see that the matrix $M^i_{ \ j}= \delta^i_{ \ j}
-\frac{8\mu g^{1/4}f^{1/4}}{n^5 \triangle_{nn}}n^in^kf_{k}$ is
non-singular at the generic points of the phase space so that the
only solution of the equation (\ref{equi}) is given by $u^i=0$ and
consequently $u^n=0$ as follows from (\ref{unui}). Then it is easy
to analyze the time evolution of the constraints $\mG_n,\mG_i$
\begin{eqnarray}
\partial_t\mG_n&=&\pb{\mG_n,H_T}=\pb{\mG_n,\bH(\bN,\bN^i)}+
\nonumber \\
&+&\int d^3\bx (v^i(\bx)\pb{\mG_n,p_i(\bx)}+v^n(\bx)
\pb{\mG_n,P_n(\bx)})=0 \ ,
\nonumber \\
\partial_t\mG_i&=&\pb{\mG_i,H_T}=\pb{\mG_i,\bH(\bN,\bN^i)}+
\nonumber \\
&+&\int d^3\bx (v^j(\bx)\pb{\mG_i,p_j(\bx)}+v^n(\bx)
\pb{\mG_i,P_n(\bx)})=0 \  \nonumber \\
\end{eqnarray}
which are four equations for four unknown functions $v^n,v^i$. Then
using the same arguments as in case of the constraints $P_n,p_i$ we
find that these equations can be solved for $v^n,v^i$ as functions
of the canonical variables and the Lagrange multipliers $\bN,\bN^i$.

Let us outline the nature of various constraints and the number of
the physical degrees of freedom in the theory. We have following
eight second class constraints: $P_n\approx 0,\ p_i\approx 0, \
\mG_n\approx 0, \ \mG_i\approx 0$.
 Solving these constraints we find
that $P_n,p_i$ vanish strongly and solving $\mG_n=0,\mG_i=0$ we can
express $n,n_i$ as functions of remaining canonical variables. We
also have  four first class constraints $P_{\bN}\approx 0,P_i\approx
0$. Gauge fixing of these constraints we can eliminate $P_{\bN},P_i$
together with $\bN,\bN^i$. Finally  we have $24$ phase space
variables $g_{ij},\pi^{ij},f_{ij},\rho^{ij}$ together with  four
first class constraints $\mR\approx 0 \ , \mR_i\approx 0$. Then
using the standard counting of the physical degrees of freedom we
find  that the number of the phase space degrees of freedom is $16$
where four of them correspond to the massless graviton while $10$ of
them can be interpreted as the massive graviton. However there are
two additional degrees of freedom corresponding to the scalar mode.
Clearly such a mode cannot be eliminated for the generic point of
the potential $\mV$. On the other hand  as we stressed in the
introduction there are examples of the suitable chosen potentials
that lead to the potentially ghost free bimetric or multimetric
theories of gravity \cite{Hinterbichler:2012cn,Hassan:2011zd}. It is
natural step to extend the analysis presented in this work to this
case as well.

 \noindent {\bf
Acknowledgement:}
 This work   was
supported by the Grant agency of the Czech republic under the grant
P201/12/G028. 

\vskip 5mm



\begin{thebibliography}{20}


\bibitem{Gourgoulhon:2007ue}
  E.~Gourgoulhon,
 \emph{``3+1 formalism and bases of numerical relativity,''}
  gr-qc/0703035 [GR-QC].


\bibitem{Arnowitt:1962hi}
  R.~L.~Arnowitt, S.~Deser, C.~W.~Misner,
 \emph{``The Dynamics of general
 relativity,''}
  [gr-qc/0405109].


\bibitem{Damour:2002ws}
  T.~Damour and I.~I.~Kogan,
 \emph{``Effective Lagrangians
 and universality classes of nonlinear bigravity,''}
  Phys.\ Rev.\ D {\bf 66} (2002) 104024
  [hep-th/0206042].


\bibitem{Hinterbichler:2012cn}
  K.~Hinterbichler and R.~A.~Rosen,
\emph{``Interacting
 Spin-2 Fields,''}
  JHEP {\bf 1207} (2012) 047
  [arXiv:1203.5783 [hep-th]].


\bibitem{Kuhnel:2012gh}
  F.~Kuhnel,
\emph{``On Instability of Certain Bi-Metric and Massive-Gravity
Theories,''}
  arXiv:1208.1764 [gr-qc].

\bibitem{Hassan:2012wr}
  S.~F.~Hassan, A.~Schmidt-May and M.~von Strauss,
\emph{``On Consistent Theories of Massive Spin-2 Fields Coupled to
Gravity,''}
  arXiv:1208.1515 [hep-th].

\bibitem{Deffayet:2012nr}
  C.~Deffayet, J.~Mourad and G.~Zahariade,
\emph{``Covariant constraints in ghost free massive gravity,''}
  arXiv:1207.6338 [hep-th].


\bibitem{Nomura:2012xr}
  K.~Nomura and J.~Soda,
\emph{``When is Multimetric
 Gravity Ghost-free?,''}
  Phys.\ Rev.\ D {\bf 86} (2012) 084052
  [arXiv:1207.3637 [hep-th]].

\bibitem{Hassan:2012wt}
  S.~F.~Hassan, A.~Schmidt-May and M.~von Strauss,
\emph{``Metric Formulation of Ghost-Free Multivielbein Theory,''}
  arXiv:1204.5202 [hep-th].

\bibitem{vonStrauss:2011mq}
  M.~von Strauss, A.~Schmidt-May, J.~Enander, E.~Mortsell and S.~F.~Hassan,
\emph{``Cosmological Solutions in Bimetric Gravity and their
Observational Tests,''}
  JCAP {\bf 1203} (2012) 042
  [arXiv:1111.1655 [gr-qc]].











\bibitem{Hassan:2011zd}
  S.~F.~Hassan and R.~A.~Rosen,
 \emph{"Bimetric Gravity
 from Ghost-free Massive Gravity,''}
  JHEP {\bf 1202} (2012) 126
  [arXiv:1109.3515 [hep-th]].

\bibitem{Banados:2011hk}
  M.~Banados, A.~Gomberoff and M.~Pino,
 \emph{``The bigravity
  black hole and its thermodynamics,''}
  Phys.\ Rev.\ D {\bf 84} (2011) 104028
  [arXiv:1105.1172 [gr-qc]].

\bibitem{Banados:2009it}
  M.~Banados and S.~Theisen,
\emph{``Three-dimensional massive gravity and the bigravity black
hole,''}
  JHEP {\bf 0911} (2009) 033
  [arXiv:0909.1163 [hep-th]].

\bibitem{Banados:2008fi}
  M.~Banados, A.~Gomberoff, D.~C.~Rodrigues and C.~Skordis,
\emph{``A Note on bigravity and dark matter,''}
  Phys.\ Rev.\ D {\bf 79} (2009) 063515
  [arXiv:0811.1270 [gr-qc]].




\bibitem{deRham:2010kj}
  C.~de Rham, G.~Gabadadze and A.~J.~Tolley,
\emph{``Resummation of Massive Gravity,''}
  Phys.\ Rev.\ Lett.\  {\bf 106} (2011) 231101
  [arXiv:1011.1232 [hep-th]].

\bibitem{Hassan:2011vm}
  S.~F.~Hassan and R.~A.~Rosen,
 \emph{"On Non-Linear Actions for Massive Gravity,''}
  JHEP {\bf 1107} (2011) 009
  [arXiv:1103.6055 [hep-th]].

\bibitem{Hassan:2011hr}
  S.~F.~Hassan and R.~A.~Rosen,
\emph{``Resolving the Ghost Problem in non-Linear Massive
Gravity,''}
  Phys.\ Rev.\ Lett.\  {\bf 108} (2012) 041101
  [arXiv:1106.3344 [hep-th]].

\bibitem{deRham:2011rn}
  C.~de Rham, G.~Gabadadze and A.~J.~Tolley,
\emph{``Ghost free
 Massive Gravity in the St\'uckelberg language,''}
  Phys.\ Lett.\ B {\bf 711} (2012) 190
  [arXiv:1107.3820 [hep-th]].

\bibitem{Hassan:2011tf}
  S.~F.~Hassan, R.~A.~Rosen and A.~Schmidt-May,
\emph{``Ghost-free Massive Gravity with a General Reference
Metric,''}
  JHEP {\bf 1202} (2012) 026
  [arXiv:1109.3230 [hep-th]].

\bibitem{Hojman:1976vp}
  S.~A.~Hojman, K.~Kuchar and C.~Teitelboim,
\emph{``Geometrodynamics Regained,''}
  Annals Phys.\  {\bf 96} (1976) 88.




\end{thebibliography}
\end{document}